\documentclass{aa}
\usepackage{graphicx}
\usepackage{txfonts}

\begin{document}

\title{Unveiling the nature of \object{IGR J16283$-$4838}\thanks{Based on
observations collected at the European Southern\- Ob\-ser\-va\-to\-ry, Chile,
under proposal ESO 075.D-0634.}}

\author{L.~J. Pellizza\inst{1,2} \and S. Chaty\inst{3} \and N.~E.
Chisari\inst{1}\thanks{Current address: Department of Astrophysical Sciences,
Princeton University, Princeton, NJ 08544.}}

\institute{Instituto de Astronom\'{\i}a y F\'{\i}sica del Espacio, C.C. 67,
Suc. 28, (1428) Buenos Aires, Argentina. \and Consejo Nacional de
Investigaciones Cient\'{\i}ficas y Te\'cnicas (CONICET), Argentina \and
Laboratoire AIM (UMR 7158 CEA/DSM-CNRS-Universit\'e Paris Diderot) Irfu/Service
d'Astrophysique, Centre de Saclay,  B\^at. 709 FR-91191 Gif-sur-Yvette Cedex,
France}

\offprints{L.J. Pellizza, \email{pellizza@iafe.uba.ar}}

\date{Received / Accepted}

\abstract{One of the most striking discoveries of the {{\em INTEGRAL}}
observatory is the existence of a previously unknown population of X-ray sources
in the inner arms of the Galaxy. The investigations of the optical/NIR
counterparts of some of them have provided evidence that they are highly
absorbed high mass X-ray binaries hosting supergiants.}{We aim to identify the
optical/NIR counterpart of one of the newly discovered {{\em INTEGRAL}} sources,
\object{IGR J16283$-$4838}, and determine the nature of this system.}{We present
optical and NIR observations of the field of \object{IGR J16283$-$4838},
and use the astrometry and photometry of the sources within it to identify its
counterpart. We obtain its NIR spectrum, and its optical/NIR spectral energy
distribution by means of broadband photometry. We search for the intrinsic
polarization of its light, and its short and long-term photometric
variability.}{We demonstrate that this source is a highly absorbed HMXB located
beyond the Galactic center, and that it may be surrounded by a variable
circumstellar medium.}{}

\maketitle

\section{Introduction}
\label{intro}

A population of new X-ray sources has been discovered by the {\em INTEGRAL}
observatory within a few tens of degrees of the direction to the Galactic center
(Negueruela \cite{Neg04}; Kuulkers \cite{Kuu05}). These sources display hard
X-ray spectra, which are usually attributed to strong absorption by dense
material lying close to the source. Their continuum spectral parameters are
typical of systems containing neutron stars or black holes. They are suspected
to be high mass X-ray binaries (HMXBs) embedded in highly absorbing media, and
for some of them massive companions were indeed identified (e.g., Filliatre \&
Chaty \cite{Fil04}; Negueruela et~al. \cite{Neg05,Neg06}; Smith et~al.
\cite{Smi06}; Masetti et~al. \cite{Mas06}; Pellizza et~al. \cite{Pel06}; Chaty
et~al. \cite{Cha08}). Investigating the nature of these sources can provide
important insight into the evolution of massive stars, the physics of compact
objects, and the mechanisms driving the accretion process.

\object{IGR J16283$-$4838} was discovered on April 7, 2005, by the {\em
INTEGRAL} observatory (Soldi et~al. \cite{Sol05}), and displayed a significant
increase in brightness over a timescale of a few days (Paizis et~al.
\cite{Pai05}). The {\em Swift} observatory performed follow-up observations on
April 13 and 15, measuring a robust position of the source (Kennea et~al.
\cite{Ken05}; Beckmann et~al. \cite{Bec05}). {\em Swift} XRT spectra are
described well by both an absorbed power law and an absorbed black body. The
X-ray absorption is high ($N_{\rm H} =
0.4$--$1.7 \times 10^{23}$~cm$^{-2}$) and variable, which suggests that it is
intrinsic to the source rather than interstellar. The absorption by the Galactic
interstellar medium in the source direction is almost an order of magnitude
lower, hence supporting the intrinsic nature of the measured source absorption.

The precise position obtained by the XRT telescope onboard {\em Swift} allowed
Rodriguez \& Paizis (\cite{Rod05}) to search for optical and NIR counterparts,
to assess the nature of the system. They found a NIR source
(2MASS~J16281083$-$4838560) only $1.7\arcsec$ from the {\em Swift} position,
with estimated magnitudes $J > 16.8$, $H > 15.8$, and $K_{\mathrm s} = 13.95 \pm
0.06$. A Magellan-Baade image in the $K$ band (Steeghs et~al. \cite{Ste05})
shows many point sources inside the {\em Swift}-XRT error circle, the position
and magnitude ($K \sim 14.1$) of the brightest one being fully consistent with
those of the 2MASS source. A mid-IR source found in the {\em Spitzer} Galactic
Legacy Midplane Survey Extraordinaire (GLIMPSE) was tentatively associated with
the {\em INTEGRAL} source by Beckmann et ~al. (\cite{Bec05}). No optical or UV
counterpart of \object{IGR J16283$-$4838} was found in {\em Swift}-UVOT data
(Beckmann et ~al. \cite{Bec05}), nor in Digitized Sky Survey (DSS) images
(Rodriguez \& Paizis \cite{Rod05}). As pointed out by Beckmann et ~al.
(\cite{Bec05}), if the association of the NIR and mid-IR sources with
\object{IGR J16283$-$4838} were correct, its spectral energy distribution would
then resemble those of high-mass X-ray binaries.

These arguments show the importance of additional multiwavelength investigations
of \object{IGR J16283$-$4838} to help identify its optical counterpart beyond
any doubt. Unveiling its properties would shed light on the nature of the new
population of X-ray sources discovered by {\em INTEGRAL}. With this aim, shortly
after the discovery of \object{IGR J16283$-$4838} we performed
optical and NIR observations of this source using the ESO
New Technology Telescope (NTT). In this paper, we present our observations
(Sect.~\ref{obs}) and results (Sect.~\ref{res}), and discuss their implications
for the nature of \object{IGR J16283$-$4838} (Sect.~\ref{disc}).

\section{Observations}
\label{obs}

Our observations of \object{IGR J16283$-$4838}, which began only 11 days after
the discovery, were carried out on the nights of 2005 April 18, 19, 21, and 28
(imaging), July 22 (imaging and polarimetry), and July 28 (spectroscopy), with
the ESO 3.5-meter NTT at La Silla Observatory, Chile. Optical and NIR images of
the field of the source were obtained with the ESO Superb-Seeing Imager 2
(SUSI2) and the Son of Isaac (SOFI) instruments respectively, as part of a
target-of-opportunity program (ESO 075.D-0634, P.~I. Chaty). NIR images were
also taken with SOFI in polarimetric mode. Low-resolution NIR spectra of the
brightest counterpart candidate proposed by Beckmann (\cite{Bec05}) were also
obtained with SOFI.

SUSI2, equipped with a mosaic of two $1024 \times 2048$ CCD detectors, was used
with $U$, $B$, $V$, $R$, and $I$ Bessel filters for optical imaging. SOFI,
equipped with a Rockwell Hg:Cd:Te $1024 \times 1024$ Hawaii array, was used with
the Large Field (LF) objective and $J$, $H$, and $K_{\mathrm{s}}$ filters for NIR
imaging. The same configuration of SOFI with the Wollaston prism, the
$K_{\mathrm s}$ filter, and the focal-plane polarimetric mask was used to obtain
NIR polarimetric images. Spectra were taken using SOFI with the LF objective and
the red and blue grisms alternatively, and their corresponding order-sorting
filters. The blue and red grisms cover the region between 0.95--1.64~$\mu$m and
1.53--2.52~$\mu$m, respectively.

On April 18 and 19, we took a set of very deep images in each filter to search
for possible counterpart candidates not detected by previous surveys, and to
perform accurate photometry of all candidates. Owing to the variable absorption,
photometric intrinsic variability was expected, hence deep NIR images were
also acquired ten days (April 28) and three months (July 22) later. A set of 306
and 234 short (2~s), contiguous exposures spanning about 2 hours each were also
taken on April 18 and 21, respectively, to search for short-term variability.
Polarimetric images were taken to search for intrinsic polarization of the NIR
light, which would be expected in the case that dust is responsible for the high
absorption in the source. We also took long exposure spectra of the candidate
proposed by Beckmann et~al. (\cite{Bec05}) to cover the whole available NIR
spectral range. Tables~\ref{img}, \ref{pol}, and \ref{spec} give the basic
parameters used for the observations. The standard reduction procedures for
optical and NIR images and NIR spectra were used to obtain the final science
images. The reduction of polarimetric images was performed following the
procedure described in the SOFI user manual using the Image Reduction and
Analysis Facility (IRAF; Tody \cite{Tod93}).

\begin{table}
\begin{center}
\caption{Optical and NIR imaging log.}
\label{img}
\begin{tabular}{cccccc}
\hline
\hline
Filter & Exposure & \# frames & Filter & Exposure & \# frames \\
 & time (s) & & & time (s) & \\
\hline
$U$ & 60 &  1 & $J$             &  60 &  27\\
$B$ & 60 &  1 & $H$             &  60 &  27\\
$V$ & 60 &  1 & $K_{\mathrm{s}}$ &  60 &  27\\
$R$ & 60 &  1 & $K_{\mathrm{s}}$ &   2 & 540\\
$I$ & 60 &  1 & \\
\hline
\end{tabular}
\end{center}
\end{table}

\begin{table}
\begin{center}
\caption{Polarimetry log.}
\label{pol}
\begin{tabular}{cccc}
\hline
\hline
Filter & Position angle & Exposure time & \# frames \\
 & ($\degr$) & (s) & \\
\hline
$K_{\mathrm{s}}$ & 0   & 60 & 6 \\
$K_{\mathrm{s}}$ & 45  & 60 & 6 \\
$K_{\mathrm{s}}$ & 90  & 60 & 6 \\
$K_{\mathrm{s}}$ & 135 & 60 & 6 \\
\hline
\end{tabular}
\end{center}
\end{table}

\begin{table}
\begin{center}
\caption{Spectroscopy log.}
\label{spec}
\begin{tabular}{lccc}
\hline
\hline
Grism & Exposure (s) & Waveband ($\mu$m) & Resolution\\
\hline
Red   & 720 & 0.95--1.64 & 930\\
Blue  & 720 & 1.53--2.52 & 980\\
\hline
\end{tabular}
\end{center}
\end{table}

\section{Data analysis and results}
\label{res}

\subsection{Astrometry}
\label{ast}

By analyzing {\em Swift} X-ray telescope (XRT) observations of {\object IGR
J16283$-$4838}, Kennea et~al. (\cite{Ken05}) obtained a preliminary position
$\mathrm{RA} = 16^{\mathrm h} 28^{\mathrm m} 10\fs7$, $\mathrm{DEC} = -48\degr
38\arcmin 55\arcsec$ ($5\arcsec$ error radius at 90\% confidence level), which
was refined later by Beckmann et~al. (\cite{Bec05}) to $\mathrm{RA} =
16^{\mathrm h} 28^{\mathrm m} 10\fs56$, $\mathrm{DEC} = -48\degr 38\arcmin
56\farcs4$ ($6\arcsec$ error radius at 90\% CL). Evans et~al. (\cite{Eva09})
developed a method to improve {\em Swift} XRT positions, reducing both
systematic and statistical uncertainties to derive error circles smaller than
$2\arcsec$ in radius. Using their web-based
form\footnote{Available at http://www.swift.ac.uk/user\_objects/} to combine all
the available observations of the source, we obtained an enhanced position for
\object{IGR J16283$-$4838}, $\mathrm{RA} = 16^{\mathrm h} 28^{\mathrm m} 10\fs97$,
$\mathrm{DEC} = -48\degr 38\arcmin 57\farcs6$ ($1.4\arcsec$ error radius at 90\%
CL), which we assume to be the most accurate estimate of the source position
available at present.

We performed the astrometry of our images to determine the position in our
frames of the {\em Swift} error circle for \object{IGR J16283$-$4838} and to
search for optical/NIR counterparts. We chose the best frame in each band for
this purpose, and selected from it a set of bright pointlike objects in
uncrowded regions. We took the coordinates of these objects from the 2MASS (for
NIR frames) and USNO B1.0 (for optical frames) catalogs (Cutri et~al.
\cite{Cut03}; Monet et~al. \cite{Mon03}), and computed the plate solution for
each frame using the IRAF {\sc ccmap} task. We obtained rms uncertainties
smaller than $0.05\arcsec$ in each coordinate for NIR frames, and $0.1\arcsec$
in each coordinate for optical frames. These uncertainties are small enough for
our purposes.

\begin{figure}
\resizebox{\hsize}{!}{\includegraphics{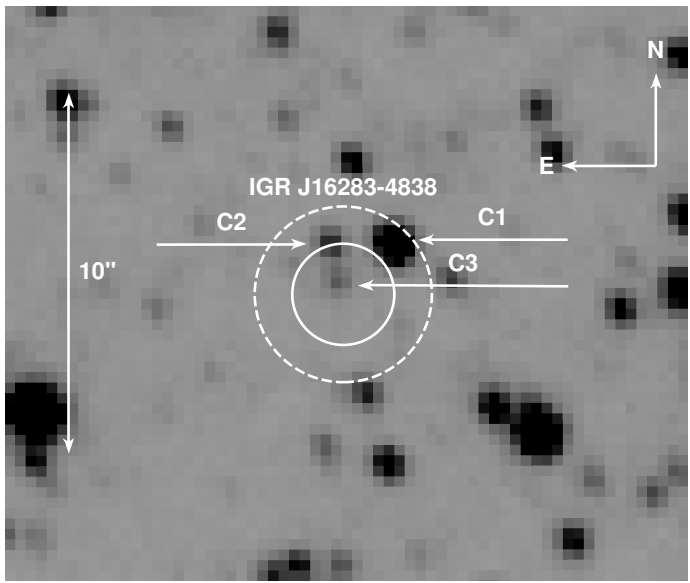}}
\caption{$K_{\mathrm s}$ band image of the field of \object{IGR J16283$-$4838}.
North is up and East is to the left. The solid-line circle is the {\em Swift}
error circle of the source ($1.4\arcsec$ radius, 90\% CL) computed using the
method of Evans et~al. (\cite{Eva09}), while the dashed-line circle is the
99.9\% CL error circle derived by us. The three counterpart candidates (C1--C3)
found inside the latter are shown in the image.}
\label{cand}
\end{figure}

In Fig.~\ref{cand}, we show the field of the source in the $K_{\mathrm{s}}$ band,
in addition to the {\em Swift}-XRT $1.4\arcsec$ error circle (solid-line
circle). Since the
confidence level of this error circle is 90\%, there is a 10\% probability that
the correct counterpart lies outside it. Hence, assuming a bivariate normal
distribution for the position error, we constructed the 99.9\% CL error circle
for the source ($2.43\arcsec$ radius, dashed-line circle in Fig.~\ref{cand}). We
assume that the probability of the source being outside this circle can be
neglected, hence we search for optical/NIR counterparts of
\object{IGR J16283$-$4838} only within the circle. As can be seen in
Fig.~\ref{cand}, three objects lie within the 99.9\% CL error circle. As pointed
out by Steeghs et~al. (\cite{Ste05}), the chance of finding an unrelated field
star within the {\em Swift}-XRT error circle is very high at this low Galactic
latitude. To illustrate this, we ran SExtractor (Bertin \& Arnouts
\cite{Ber96}) on our full $K_{\mathrm{s}}$ frame using a detection threshold of
$3 \sigma$ above background, and found 11\,107 sources within its $5.9\arcmin
\times 5.9\arcmin$ field. This implies an expected mean of 1.64 objects inside
the 99.9\% CL error circle, which is consistent with the three we found (C1--C3,
their positions are given in Table~\ref{posc}). The brightest candidate (C1 =
\object{2MASS J16281083$-$4838560}) is located at $2\arcsec$ from the nominal
position of the X-ray source, and the other two candidates are closer to it, but
several magnitudes fainter. None of the candidates is seen in any of the optical
images, which is consistent with the strong absorption towards this region ($l
= 335.3\degr$, $b = +0.1\degr$, $A_V = 12.3$~mag; Beckmann et~al. \cite{Bec05}).
At this point, there is no preferred candidate: astrometry alone cannot help us
identify the counterpart of \object{IGR J16283$-$4838}, until subarcsecond
positioning such as that achievable with {\em Chandra} is available for this
source. A $\log N$-$\log S$ diagram of the sources in our $K_{\mathrm s}$ frame
shows that there is a 14\% probability of finding at least one unrelated NIR
source brighter than C1 within the 99.9\% CL error, which increases to 55\% and
70\% for C2 and C3, respectively. This result indicates that C1 is the strongest
counterpart candidate.

\begin{table}
\begin{center}
\caption{Positions of NIR counterpart candidates of \object{IGR J16283$-$4838},
and their distances to the nominal position of the X-ray source.}
\label{posc}
\begin{tabular}{lccc}
\hline
\hline
Id. & $\alpha$ (J2000) & $\delta$ (J2000) & Distance to X source\\
 & (h\,m\,s) & ($\degr\,\arcmin\,\arcsec$) & ($\arcsec$) \\
\hline
C1 & 16:28:10.83 & $-$48:38:56.2 & 2.0 \\
C2 & 16:28:11.01 & $-$48:38:56.3 & 1.4 \\
C3 & 16:28:10.99 & $-$48:38:57.3 & 0.4 \\
\hline
\end{tabular}
\end{center}
\end{table}

\subsection{Spectroscopy}
\label{spe}

\begin{figure}
\resizebox{\hsize}{!}{\includegraphics{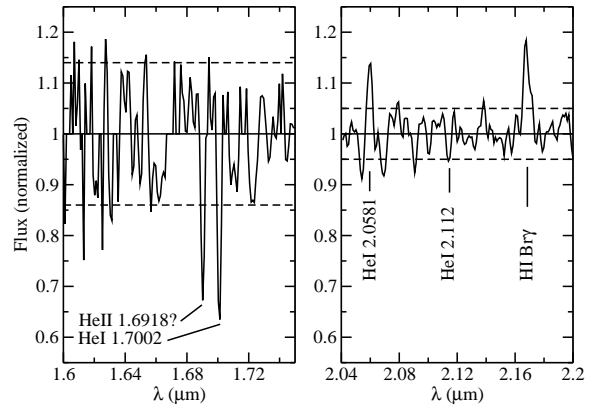}}
\caption{NIR spectrum of candidate C1 in the ranges 1.6--1.75~$\mu$m (left
panel) and 2.04--2.2~$\mu$m (right panel) showing the identified lines, among
which we found \ion{H}{i} and \ion{He}{i} lines typical of early-type stars. The
horizontal solid line delineates the position of the continuum, while the dashed
lines indicate the 2$\sigma$ uncertainties. The rest of the spectrum (see
Table~\ref{spec}) is not shown because its faintness prevented us from
extracting useful data.}
\label{spectra}
\end{figure}

The faintness of all counterpart candidates makes it difficult to obtain their
spectra, C1 being the only case in which we succeeded. Given this situation, a
careful extraction of the spectrum was performed, especially in terms of sky
subtraction and the removal of telluric features. In Fig.~\ref{spectra}, we
present the NIR spectra of this object taken with the red grism of SOFI (the
blue grism spectrum shows a large absorption, which renders it useless). As can
be seen in Fig.~\ref{spectra}, the uncertainty in the continuum position is
3--8\%, allowing us to identify only a few features in it, which are listed in
Table~\ref{lines}.

The \ion{H}{i} Br$\gamma$ line at 2.1655~$\mu$m is clearly seen in emission.
Other hydrogen lines are not detected, probably because their intensities
would be below the continuum noise level. In the spectrum of C1, the \ion{He}{i}
1.7002~$\mu$m and 2.0581~$\mu$m lines are also present, visible in absorption
and emission, respectively. The 2.112/2.113~$\mu$m \ion{He}{i} line is
marginally detected at the 2$\sigma$ level. Since hydrogen and helium lines are
the main spectral features, some in emission, C1 may be an early-type star. We
also searched for ionized helium features. According to Hanson et~al.
(\cite{Han05}), the strongest \ion{He}{ii} line in this region of the spectrum
is at 2.1885~$\mu$m, but we do not detect it. An absorption feature clearly
detected near 1.6918~$\mu$m in our spectrum could be attributed to the
corresponding \ion{He}{ii} line. However, that this line should be comparable
to or weaker than the 2.1885~$\mu$m one (Hanson et~al. \cite{Han05}) suggests
that it might be an artifact due to subtraction of the sky lines. The lack of
\ion{He}{ii} lines then implies that C1 is a late O-type or early B-type star.
Apart from hydrogen and helium lines, we marginally detect other absorption
features, which we cannot identify.

The clear evidence of Br$\gamma$ in emission is particularly interesting. Hanson
et~al. (\cite{Han05}) found this behavior only in supergiants. The stars in
their sample with Br$\gamma$ in emission are of luminosity class I. Among them,
early-O supergiants show a strong and broad line, late-O and early-B supergiants
show a weaker and narrower line, while supergiants later than B1 always show
this line in absorption. As Br$\gamma$ in our spectrum is relatively weak, C1
seems to be a late-O or early-B supergiant, consistent with the absence of
ionized helium lines. Hence, we propose that the spectrum provides strong
evidence that C1 is a late-O or early-B supergiant.

Stars such as C1 are rare objects. The catalog of Reed (\cite{Ree03}) lists
$\sim$6000 OB supergiants in the solar neighborhood, and approximately the same
number of OB dwarfs. Reed (\cite{Ree05}) extrapolates the number of OB dwarfs in
his catalog to obtain a total number of $\sim 1.5-2.5 \times 10^5$ in the
Galaxy. Hence, as a rough estimate, the same values could be assumed for the
number of OB supergiants in the Galaxy. Using these values and assuming OB
supergiants to be uniformly distributed in the sky within the range $-10\degr <
b < 10\degr$, the probability of finding an unrelated OB supergiant within our
99.9\% CL error circle is $2-3 \times 10^{-5}$. For stars such as C1, clearly a
subset of OB supergiants, this probability must be even lower. This result
strongly suggests that C1 is indeed the correct counterpart to \object{IGR
J16283$-$4838}.

\begin{table}
\begin{center}
\caption{Lines identified in the spectrum of C1.}
\label{lines}
\begin{tabular}{cccl}
\hline
\hline
$\lambda_{\mathrm{obs}}$ & $\lambda_{\mathrm{lab}}$ & EW & Species \\
($\mu$m) & ($\mu$m) & ({\AA}) \\
\hline
$1.690 \pm 0.003$ & 1.6918?           &   $8 \pm 2$ & \ion{He}{ii}? \\
$1.701 \pm 0.003$ & 1.7002            &  $10 \pm 2$ & \ion{He}{i} \\
$2.060 \pm 0.003$ & 2.0581            &  $-6 \pm 1$ & \ion{He}{i} \\
$2.114 \pm 0.003$ & 2.1120/2.1132     &   $2 \pm 1$ & \ion{He}{i} \\
$2.168 \pm 0.003$ & 2.1655            & $-10 \pm 1$ & \ion{H}{i} (Br$\gamma$) \\
\hline
\end{tabular}
\end{center}
\end{table}

\subsection{Photometry}
\label{phot}

Aperture photometry of all the candidates on the frames of April 28 was
performed using the IRAF package {\sc apphot}. This date was chosen because we
have standard star observations for this night to calibrate our frames. For each
frame, we measured the instrumental optical and NIR magnitudes of objects C1--C3
(or their lower limits) in the $UBVRI$ and $JHK_{\mathrm s}$ bands. The zero
point was directly obtained from measurements of standard stars at the same
airmass values. None of the counterpart candidates was detected in any of the
optical bands, the upper limits to their magnitudes being $U > 19.4$, $B >
20.6$, $V > 20.5$, $R > 20.0$, and $I > 17.6$. However, we detect almost all of
them in the NIR bands. Their NIR magnitudes are shown in Table~\ref{magc}. We
note that our $K_{\mathrm s}$ magnitude of C1 is in very good agreement with that
measured by Steeghs et~al. (\cite{Ste05}).

\begin{table}
\begin{center}
\caption{Optical and NIR magnitudes of the counterpart candidates of the high
energy source \object{IGR J16283$-$4838}.}
\label{magc}
\begin{tabular}{lccc}
\hline
\hline
Id. & $J$ & $H$ & $K_\mathrm{s}$ \\
 & $\pm 0.09$~mag & $\pm 0.04$~mag & $\pm 0.04$~mag \\
\hline
C1 &   19.14 &   15.75 &   14.05 \\
C2 &   20.47 &   15.55 &   16.27 \\
C3 & $>$20.5 &   17.15 &   15.60 \\
\hline
\end{tabular}
\end{center}
\end{table}

\begin{figure}
\resizebox{\hsize}{!}{\includegraphics{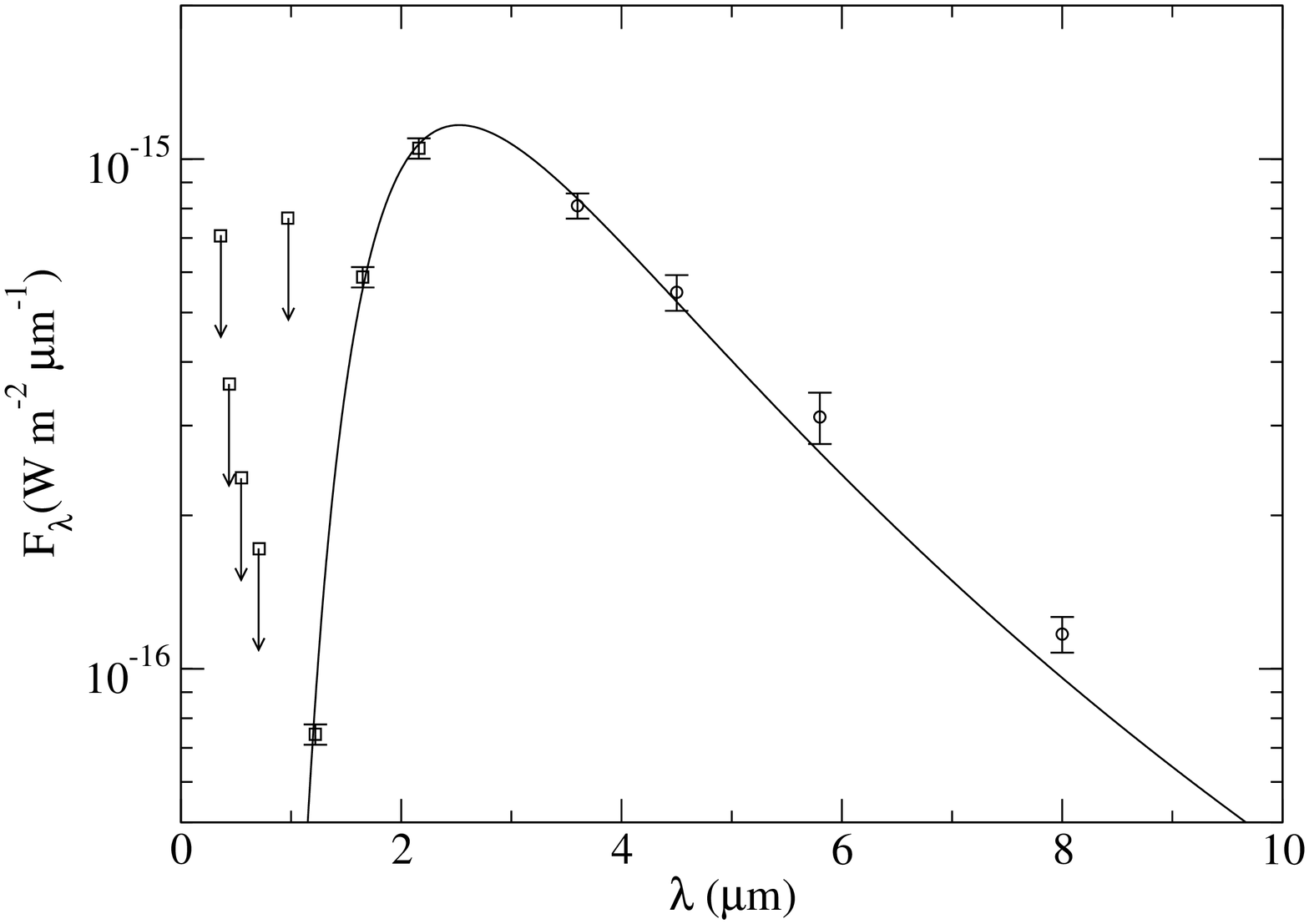}}
\caption{Spectral energy distribution $F_\lambda (\lambda)$ of C1. Squares
represent the fluxes observed by us and circles those from GLIMPSE data. The
solid line shows the best-fit of an absorbed black body emitting at the
effective temperature of a late O or early B star ($T = 30\,000$~K), with an
absolute magnitude $M_V = -6.3$.}
\label{sed}
\end{figure}

Figure~\ref{sed} shows the observed spectral energy distribution (SED) of C1.
Since the position of the GLIMPSE source \object{SSTGLMC G335.3268+00.1016} is
consistent with that of \object{2MASS J16281083$-$4838560} = C1, its photometry
was included in the SED. Figure~\ref{sed} shows that the fluxes of both sources
are on the same order of magnitude, suggesting that a single model could fit
both observations.

At mid-IR wavelengths, the SED shows a decreasing tail resembling that of a high
temperature black body, while in the NIR it has a sharp cut, most probably due
to absorption. Given that in the mid-IR the extinction is negligible, a distance
for the source can be derived, assuming a luminosity and spectrum for the
source. On the other hand, the sharp cut in the NIR can help us determine the
value of the total visual extinction. We performed an absorbed black-body fit to
the observed SED to constrain these properties. Following our spectral analysis,
we assumed that the emission comes from a blue supergiant of absolute visual
magnitude $M_V = -6.3$ (Martins et~al. \cite{Mar05}) at an unknown distance $d$,
which emits like a 30\,000~K black body. The radiation is absorbed by a medium
of unknown visual extinction $A_V$, which is assumed to follow the extinction
law proposed by Cardelli et~al. (\cite{Car89}). Best-fit values obtained are $d
= 17.2$~kpc and $A_V = 28.9$ magnitudes. Martins et~al. (\cite{Mar05}) find a
dispersion of $\sim$0.5~mag in the absolute magnitudes of blue supergiants,
which translates into a distance range of 13.6--21.6~kpc for the source. The
value of $A_V$ shows no change with the assumed value of $M_V$. The distance and
visual extinction are almost insensitive to a change in the assumed temperature
for the star. Varying the latter in the range 25\,000--35\,000~K produces a
change of $<0.8$~kpc in distance and $<0.1$~mag in $A_V$. The extinction can be
transformed into a hydrogen column density of $0.52 \times 10^{23}$~cm$^{-2}$,
in excellent agreement with the value found by Beckmann et~al. (\cite{Bec05})
for the source in quiescence. That the SED of C1 and the GLIMPSE source
can be fitted by a single absorbed black-body model, of a temperature and
luminosity consistent with our spectra of C1, which has an absorption in full
agreement with that measured by Beckmann et~al. \cite{Bec05} for
\object{IGR J16283$-$4838}, strongly supports the suggestion that the three
types of emission come from the same object. Our results indicate that this
object is a highly absorbed high-mass X-ray binary with a blue supergiant
secondary, located beyond the Galactic center but still inside the Galaxy.

\subsection{Variability}
\label{lc}

\begin{figure}
\resizebox{\hsize}{!}{\includegraphics{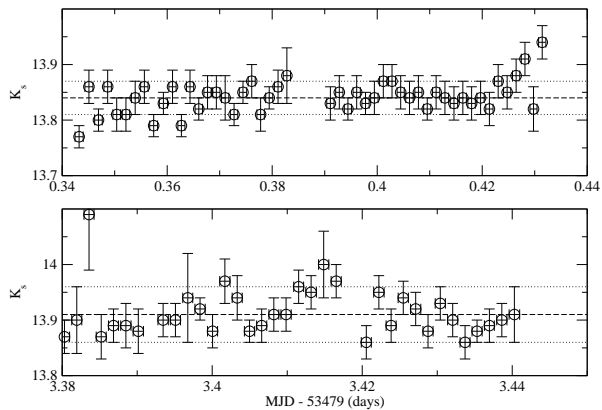}}
\caption{Short-term NIR light curve of C1. Abscissas are given as modified
Julian dates (MJD) since April 19, 2005, 0$^h$ UT. The dashed lines represent
the mean magnitude of the source, while the dotted lines show the standard
deviation in the data about the mean. C1 shows irregular variations
statistically consistent with the photometric errors.}
\label{curves}
\end{figure}

To search for the short-term variability of C1, we used the 540
$K_{\mathrm s}$-band short exposures (see Table~\ref{img}) to construct two light
curves, one spanning $\sim$7~ks on April 18, and the other $\sim$5~ks on April
21. The images were averaged in groups of 6 to increase the accuracy of the
photometry. To avoid the effect of the variable airmass and sky transparency in
the NIR, we performed differential aperture photometry of C1 against three
comparison stars in its field. The instrumental magnitude difference between the
comparison stars was found to be constant within the photometric errors, hence
to the accuracy of our photometry ($\sim$0.02~mag) these stars can be considered
non-variable. Their magnitudes were calibrated against a standard star observed
at the same airmass as one of the frames. In Fig.~\ref{curves}, we present the
light curves obtained. Both curves exhibit random variations that are fully
consistent with photometric uncertainties. A Lomb-Scargle periodogram does not
display any periodicity, indicating that the data are consistent with a constant
source.

\begin{figure}
\resizebox{\hsize}{!}{\includegraphics{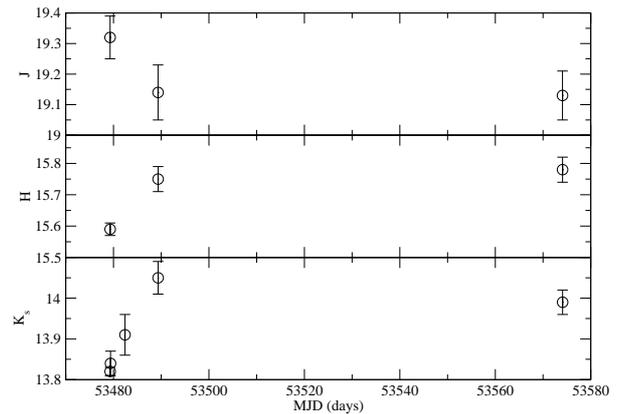}}
\caption{Long-term NIR light curve of C1. Abscissas are given as modified Julian
dates (MJD). C1 shows a steady decline in brightness in the $H$ and
$K_{\mathrm s}$ bands during late-April, but the opposite behavior in the $J$
band. The July data imply that the source reached a steady state.}
\label{longcurve}
\end{figure}

We also searched for long-term variability, using the observations on April 18,
21, and 28 and July 22 to monitor possible magnitude changes on scales of days
or months. Figure~\ref{longcurve} shows that these changes do occur in C1, which
was brighter in the NIR in mid April, fading later by $\sim$0.2~mag in both $H$
and $K_{\mathrm s}$. The $J$ band data present the opposite behavior, displaying
a marginal brightening in this band. None of the other two counterpart
candidates (C2--C3) shows magnitude variations between April and July
2005. The long-term variability of C1 in the NIR bands is interesting, and
in addition to the properties discussed in former sections of the present work,
makes a strong case for C1=\object{2MASS J16281083$-$4838560}, \object{SSTGLMC
G335.3268+00.1016}, and \object{IGR J16283$-$4838} being different emissions of
the same astrophysical system.

\subsection{Polarimetry}
\label{polar}

NIR polarimetry in the $K_{\mathrm s}$ band was performed using the observations
taken with the Wollaston prism at different position angles. Images were reduced
using the procedure described in the SOFI user manual. Aperture photometry was
performed for candidate C1 and several stars in the field to determine their
fluxes. Standard unpolarized (HD 125184) and polarized stars (HD 150193, $P =
1.68 \pm 0.02\%$ at a position angle of $60 \pm 1\degr$) were also observed to
ensure that instrumental and sky polarization were appropriately removed. The
measurement of field stars allowed us to estimate the foreground interstellar
polarization, obtaining a value of $P_{\mathrm f} = 3.7 \pm 0.5\%$ at a position
angle (north through east) of $33 \pm 1\degr$. A deviation of C1 polarization
from this value would indicate an intrinsic polarization of the source. For C1,
we obtained $P = 4.6 \pm 1.0\%$ at a position angle of $35 \pm
1\degr$, statistically consistent with the foreground value. The intrinsic
polarization of the light of C1, if any, must be lower than 1\%. Hence, our
data suggest that the source is not polarized, which is consistent with stellar
light emitted isotropically from a supergiant star.

\section{Discussion}
\label{disc}

Using optical and NIR images, spectra and polarimetry, we have searched for the
counterpart of the high-energy source \object{IGR J16283$-$4838}. We have found
three candidates inside the {\em Swift} error circle of the source, which is
consistent with the expected number of chance superpositions. However, several
pieces of evidence suggest that the brightest one (\object{2MASS
J16281083$-$4838560}) is the correct counterpart to the high energy source. This
source, coincident with a MIR GLIMPSE source (\object{SSTGLMC
G335.3268+00.1016}) exhibits NIR spectral features that are typical of late-O or
early-B supergiant stars. Moreover, the combined NIR--MIR SED of the 2MASS and
GLIMPSE sources is most closely fitted by an absorbed black body of luminosity
and temperature consistent with these stellar types, and with an absorption in
full agreement with that measured in X-rays for \object{IGR J16283$-$4838}. The
2MASS source also displayed a steady luminosity decrease during April 18--28,
2005, a few days after the X-ray source flare of April 7--10 and variations in
its absorption during April 14--15. Hence, we infer that our results imply that
\object{2MASS J16281083$-$4838560}, \object{SSTGLMC G335.3268+00.1016}, and
\object{IGR J16283$-$4838} are the same source.

Given the association discussed above, our data strongly imply that \object{IGR
J16283$-$4838} is a Galactic high-mass X-ray binary, as suspected by Beckmann
et~al. (\cite{Bec05}). Our classification of the donor in this system as a
late-O or early-B blue supergiant, in addition to our photometry displaying a
SED consistent with a high-mass stellar source in the Galaxy, with an
extinction close to that derived from X-ray observations, makes the
classification of \object{IGR J16283$-$4838} highly secure. Our data cannot help
us determine the nature of the accretor, however, as discussed by Beckmann
et~al. (\cite{Bec05}), this source being similar to those in the class of
highly-absorbed HMXBs discovered by {\em INTEGRAL}, such as \object{IGR
J16318$-$4848} (Filliatre \& Chaty \cite{Fil04}, see also Chaty et~al.
\cite{Cha08}) or \object{IGR J19140+0951} (Rodriguez et~al. \cite{Rdr05}). Most
systems in this class have been shown to contain neutron stars as accretors. We
note that our distance estimate changes the value of the source luminosity
considerably, with respect to that estimated by Beckmann et~al. (\cite{Bec05}).
For our 13.6--21.6~kpc range, the luminosity during the flare would have been
in the range $L = 10^{36.8-37.2}\,\mathrm{erg \, s}^{-1}$, while in quiescence it
would have been $L = 10^{35.5-35.9}\,\mathrm{erg \, s}^{-1}$. These values are
far higher than those derived by Beckmann et~al. (\cite{Bec05}), but still
consistent with a neutron star accretor. However, the possibility of a black
hole accretor cannot be ruled out by the available data.

The origin of the absorption in this system is another interesting point. The
variability in the X-ray absorption implies that this is caused by a
circumstellar medium. A strong, variable stellar wind from the early-type donor
would be a natural explanation, as pointed out by Beckmann et~al.
(\cite{Bec05}). Our results suggest that the NIR and X-ray absorption are
related, since the absorption obtained from our fit to the NIR--MIR SED ($A_V =
28.9$ magnitudes on April 28, 2005) translates into a hydrogen column density of
$0.52 \times 10^{23}$~cm$^{-2}$, similar to that obtained by Beckmann et~al.
(\cite{Bec05}) by fitting the X-ray spectra of the source on April 13 and April
15, 2005 (of 0.6 and $0.4 \times 10^{23}$~cm$^{-2}$, respectively). This would
point to an extended medium in which both the primary and the donor would be
embedded. Both the X-ray absorption and NIR luminosity interestingly exhibit
variations on timescales of days. The latter could be attributed within this
picture to absorption variations, although no simultaneous X-ray and NIR data
are available to test this suggestion by searching for correlations between both
absorption values.

If this picture is correct, the contrasting behavior of the $J$-band luminosity
on the one hand, and $H$--$K_{\mathrm s}$ luminosity on the other remains
intriguing. Emission at longer wavelengths (MIR--FIR), correlated with
NIR/X-ray absorption, that has a tail reaching the NIR, would be a possible
explanation of this behavior. In this case, the $J$-band variations may be
controlled by the absorption of the system, while those in $H$ and
$K_{\mathrm s}$ bands could be more strongly dependent on the emission. Warm
circumstellar dust in the absorber medium that reprocesses the absorbed light
may produce the proposed MIR--FIR radiation. Unfortunately our data are
insufficient to perform a quantitative test of this hypothesis. Follow-up
observations of the flare and post-flare behavior of the source in the NIR and
MIR ranges, achievable for example with instruments such as SOFI at NTT and
VISIR at VLT (e.g., Rahoui et~al. \cite{Rah08}) would be interesting to address
this point. Undoubtedly, additional observations of this source are needed to
reveal the particular mechanism responsible for the accretion in this class of
high-mass X-ray binaries.

\begin{acknowledgements}

We acknowledge the comments and suggestions by the anonymous referee, which
greatly enhanced our manuscript.
This publication makes use of data products from the Two Micron All Sky Survey,
which is a joint project of the University of Massachusetts and the Infrared
Processing and Analysis Center / California Institute of Technology, funded by
the National Aeronautics and Space Administration and the National Science
Foundation. This research has made use of the SIMBAD database and VizieR Service
operated at CDS, Strasbourg, France, and of NASA's Astrophysics Data System
Bibliographic Services. This work was supported by the Centre National d'Etudes
Spatiales (CNES). It is based on observations obtained through MINE: the
Multi-wavelength INTEGRAL Network. LJP acknowledges support by grant PICT
2007-00848 of Argentine ANPCyT.

\end{acknowledgements}


\begin{thebibliography}{}

\bibitem[2005]{Bec05}
Beckmann, V., Kennea, J.~A., Markwardt, C., et~al. 2005,
\apj, 631, 506

\bibitem[1996]{Ber96}
Bertin, E., \& Arnouts, S. 1996,
\aaps, 117, 393

\bibitem[1989]{Car89}
Cardelli, J.A, Clayton, G.C., \& Mathis, J.S., 1989,
\apj, 345, 245

\bibitem[2008]{Cha08}
Chaty, S., Rahoui, F., Foellmi, C., et~al. 2008,
\aap, 484, 783

\bibitem[2003]{Cut03}
Cutri, R.~M., Skrutskie, M.~F., van Dyk, S., et al. 2003,
2MASS All-Sky Catalog of Point Sources, University of Massachusetts and
Infrared Processing and Analysis Center, (IPAC / California Institute of
Technology). Vizier online catalog II/246.

\bibitem[2009]{Eva09}
Evans, P.~A., Beardmore, A.P., Page, K.~L., et al. 2009,
\mnras, 397, 1177

\bibitem[2004]{Fil04}
Filliatre, P., \& Chaty, S. 2004,
\apj, 616, 469

\bibitem[2005]{Han05}
Hanson, M.~M., Kudritzki, R.-P., Kenworthy, M.~A., et al. 2005,
\apjs, 161, 154

\bibitem[2005]{Ken05}
Kennea, J.~A., Burrows, D.~N., Nousek, J.~A., et al. 2005,
ATel, 459

\bibitem[2005]{Kuu05}
Kuulkers, E. 2005,
AIP Conference Proceedings, 797, 402

\bibitem[2005]{Mar05}
Martins, F., Schaerer, D., \& Hillier, D.~J. 2005,
\aap, 436, 1049

\bibitem[2006]{Mas06}
Masetti, N., Pretorius, M.~L., Palazzi, E., et~al. 2006,
\aap, 449, 1139

\bibitem[2003]{Mon03}
Monet, D.~G., Levine, S.~E., Canzian, B., et~al. 2003,
\aj, 125, 984

\bibitem[2004]{Neg04}
Negueruela, I. 2004,
in ``The Many Scales of the Universe - JENAM 2004 Astrophysics Reviews'', eds.
J.~C. del Toro Iniesta, et~al., Proc of the Joint European and Spanish
Astronomical Meeting, Granada, Spain, September 2004, arXiv:astro-ph/0411759

\bibitem[2005]{Neg05}
Negueruela, I., Smith, D.~M., \& Chaty, S. 2005,
ATel, 429

\bibitem[2006]{Neg06}
Negueruela, I., Smith, D.~M., Harrison, Th.~E., \& Torrej\'on, J.~M. 2006,
\apj, 638, 982

\bibitem[2005]{Pai05}
Paizis, A., Miller, J.~M., Soldi, S., \& Mowlavi, N. 2005,
ATel, 458

\bibitem[2006]{Pel06}
Pellizza, L.~J., Chaty, S., \& Negueruela, I. 2006,
\aap, 455, 653

\bibitem[2008]{Rah08}
Rahoui, F., Chaty, S., Lagage, P.-O., \& Pantin, E. 2008,
\aap, 484, 801

\bibitem[2003]{Ree03}
Reed, B.~C. 2003,
\aj, 125, 2531

\bibitem[2005]{Ree05}
Reed, B.~C. 2005,
\aj, 130, 1652

\bibitem[2005]{Rdr05}
Rodriguez, J., Cabanac, C., Hannikainen, D.~C., et~al 2005,
\aap, 432, 235

\bibitem[2005]{Rod05}
Rodriguez, J., \& Paizis, A. 2005,
ATel, 460

\bibitem[2006]{Smi06}
Smith, D.~M. et~al. 2006,
\apj, 638, 974

\bibitem[2005]{Sol05}
Soldi, S., Brandt, S., Domingo Garau, A., et~al. 2005,
ATel, 456

\bibitem[2005]{Ste05}
Steeghs, D., Torres, M.~A.~P., Jonker, P.~G., et~al. 2005,
ATel, 478

\bibitem[1993]{Tod93}
Tody, D. 1993,
in ``Astronomical Data Analysis Software and Systems II'', A.S.P. Conference
Ser., Vol 52, eds. R.~J. Hanisch, R.~J.~V. Brissenden, \& J. Barnes, 173

\end{thebibliography}
\end{document}